\documentclass[preprint2]{aastex}

\slugcomment{accepted by ApJ}

\shorttitle{Galactic cosmic rays with TRACER}
\shortauthors{Obermeier et al.}

\begin{document}

\title{Energy spectra of primary and secondary cosmic-ray nuclei measured with TRACER}

\author{A. Obermeier\altaffilmark{1,2}, M. Ave\altaffilmark{1,3}, P. Boyle\altaffilmark{1,4}, Ch. H\"oppner\altaffilmark{1,5}, J. H\"orandel\altaffilmark{2} and D. M\"uller\altaffilmark{1}}

\email{a.obermeier@astro.ru.nl}

\altaffiltext{1}{University of Chicago, Chicago, IL 60637, USA}
\altaffiltext{2}{Radboud Universiteit Nijmegen, 6525 HP Nijmegen, The Netherlands}
\altaffiltext{3}{Now at Karlsruhe Institute of Technology (KIT), Germany}
\altaffiltext{4}{Now at McGill University, Montreal, Canada}
\altaffiltext{5}{Now at Technische Universit\"at M\"unchen, Germany}

\begin{abstract}
The TRACER cosmic-ray detector, first flown on long-duration balloon (LDB) in 2003 for observations of the major primary cosmic-ray nuclei from oxygen ($Z=8$) to iron ($Z=26$), has been upgraded to also measure the energies of the lighter nuclei, including the secondary species boron ($Z=5$). The instrument was used in another LDB flight in 2006. The properties and performance of the modified detector system are described, and the analysis of the data from the 2006 flight is discussed. The energy spectra of the primary nuclei carbon ($Z=6$), oxygen, and iron over the range from 1~GeV~amu$^{-1}$ to 2~TeV~amu$^{-1}$ are reported. The data for oxygen and iron are found to be in good agreement with the results of the previous TRACER flight. The measurement of the energy spectrum of boron also extends into the TeV~amu$^{-1}$ region. The relative abundances of the primary nuclei, such as carbon, oxygen, and iron, above $\sim10$~GeV~amu$^{-1}$ are independent of energy, while the boron abundance, i.\ e.\ the B/C abundance ratio, decreases with energy as expected. However, there is an indication that the previously reported $E^{-0.6}$ dependence of the B/C ratio does not continue to the highest energies.
\end{abstract}


\section{Introduction}

Direct measurements of the energy spectra of cosmic ray nuclei that extend into the high-energy region, i.e. energies around and above $10^{12}$~eV~amu$^{-1}$,  require the exposure of large detectors over extended observation times. Only one such measurement has thus far been realized in space flight, the Cosmic-Ray Nuclei detector (CRN) that was flown on the Space Shuttle~\citep{CRNdetector}. Long-duration balloon (LDB) flights permit exposures of the order of weeks, and several LDB instruments have been developed: TRACER (“Transition Radiation Array for Cosmic Energetic Radiation”)~\citep{NIM}, which is the subject of this paper, and two instruments that use short calorimeters for the energy measurement, ATIC~\citep{ATICdetector} and CREAM~\citep{CREAMdetector}. TRACER was designed and constructed at the University of Chicago, utilizing the heritage of CRN.
In contrast to the calorimetric measurements, a nuclear interaction is not required to occur, because charge and energy of cosmic rays are determined exclusively via electro-magnetic interactions. This allows TRACER to employ light-weight materials that consequently allow the instrument to be very large. It has a geometric factor of $\sim$5 m$^2$ sr (detector area $2\times2$~m$^2$) and is currently the largest balloon-borne cosmic-ray detector. 

TRACER  had a two-week circum-polar LDB flight in Antarctica in 2003 (LDB1). The results from this flight~\citep{AveMeas} provide a comprehensive set of measurements of the individual energy spectra of the major primary cosmic-ray nuclei from oxygen ($Z=8$) to iron ($Z=26$), covering the energy range from $\sim1$~GeV~amu$^{-1}$ to several TeV~amu$^{-1}$. In total energy, the measurement reached energies well above 10$^{14}$~eV for the more abundant species. The results from this flight indicate that power-law fits in energy to the measured spectra can be made above 20~GeV~amu$^{-1}$ with the same power-law exponent,  $\alpha=2.67\pm0.05$, for all elemental species. There seems to be no noticeable dependence of this exponent on nuclear charge Z. This result appears to be surprising, and in fact, an analysis of the TRACER results in the context of simple propagation and source models of cosmic rays~\citep{AveInt} indicates that the observed spectra might not be exact power-laws but exhibit a slight concave curvature due to the competition of  interaction losses and energy-dependent diffusive escape during the propagation through the Galaxy. If this curvature existed, it could be obscured by the statistical uncertainties in the measurement.

Within measurement uncertainties, the TRACER data appear to be consistent with the results reported from ATIC and CREAM. In particular, the CREAM group also has reported a common power law exponent of $2.66\pm0.04$ for the measured spectra~\citep{CREAMspectra}. More recently, the CREAM group has claimed a ``discrepant'' upturn in the measured spectra at $\sim200$~GeV~amu$^{-1}$~\citep{DiscrHard} but again, the statistical significance is quite limited.

The cosmic-ray intensities measured near the Earth are significantly affected by Galactic propagation and do not directly reflect the characteristics of the cosmic-ray sources.  At low energies ($<$~1~GeV~amu$^{-1}$), the time for diffusive propagation through the Galaxy is $\sim1.5\cdot10^7$ years, at an interstellar path length of a few g~cm$^{-2}$~\citep{garcia,yanasak}. It has been known for a long time that the path length decreases at relativistic energies, apparently as a power law proportional to $E^{-0.6}$~\citep{Juliusson,yanasak,HEAObc,CRNbc}. Whether this strong decrease persists to the highest energies, is an open question~\citep{AveInt}.  While the particles diffuse through interstellar space, they can also be lost by spallation reactions with the components of the interstellar gas. The spallation cross sections depend on the mass number $A$ of the projectile approximately like $A^{2/3}$, hence heavy nuclei are more strongly affected than light ones. In order to determine the characteristics of the cosmic-ray sources, these two propagation effects, one depending on energy and the other on nuclear mass number, must be understood to deconvolve the measured spectra.

The propagation path length of cosmic rays can naturally be determined by measuring the intensities and energy spectra of the secondary nuclei, i.e. those elements that are produced exclusively by interstellar spallation. This group includes the light elements lithium, beryllium, and boron, where boron is the most studied element. At relativistic energies, quite accurate measurements exist from HEAO~\citep{HEAObc}, but do not extend beyond $\sim30$~GeV~amu$^{-1}$. At higher energies, up to a few hundred GeV~amu$^{-1}$ the boron intensity has been measured by CRN~\citep{CRNbc} and results on the relative abundance of boron have been reported by CREAM~\citep{CREAMbc}, but the data are severely limited by statistical uncertainties at the highest energies. 

The TRACER instrument could not include an energy measurement of the light nuclei below oxygen in LDB1 because of limitations in the dynamic range of its detector electronics. A major upgrade of TRACER was performed after that flight. The instrument is now sensitive to  the entire charge range $4<Z<28$ and was launched on an LDB flight from Kiruna, Sweden in 2006 (LDB2). In the following, we will briefly describe the modified detector system, discuss the data analysis, and present the results from the this flight, as well as the combined data set from LDB1 and LDB2.

\section{The Measurement}

\subsection{The Detector and Measurement Principle}
\label{sec:principle}

The TRACER instrument represents a combination  of electronic detectors, comprising plastic scintillator sheets, acrylic \v{C}erenkov counters, and arrays of single-wire proportional tubes. The overall dimensions of the detector system are quite large, leading to an aperture of 4.73~m$^2$~sr. The instrument and the measurement process for LDB1 have been described in some detail by \citet{AveMeas}, and a broader, technical description which also includes the modifications for LDB2, is being published \citep{NIM}. Therefore, we will here only briefly review the measurement process and emphasize some aspects which are of specific importance for the 2006 measurement.

\begin{figure}[tb]
 \includegraphics[width=\linewidth]{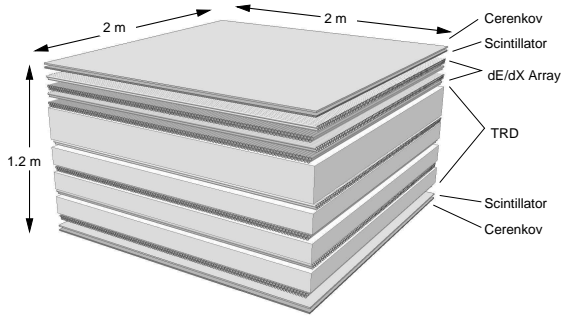}
 \caption{Schematic view of the detector elements of TRACER.\label{fig1}}
\end{figure}

A schematic drawing of TRACER is shown in Figure~\ref{fig1}. There are two identical pairs of scintillator and \v{C}erenkov counters on top and bottom of the instrument, respectively, with each counter 200~cm x 200~cm in area. Each counter is viewed by 24 photomultiplier tubes (PMTs) via wavelength-shifter bars. The scintillators provide the coincidence trigger for the measurement.

The space between these counter-pairs is occupied by arrays of gas-filled, single-wire proportional tubes and layers of plastic-fiber radiators. There is a total of 1584 tubes, each 2~cm in diameter and 200~cm in length. They are arranged in double-layers in densest packing and are alternately oriented in two orthogonal directions. Thus, there is a total of 8 double layers of $2\times99$ tubes, each. The four upper double layers measure the ionization energy loss (``dE/dx''). Each of the four remaining layers is located below a radiator layer and measures ionization loss with transition radiation superimposed (``dE/dx + TR'').  The arrangement and packing of the fiber radiators is identical to that used for the CRN instrument~\citep{CRNdetector}, and for LDB1~\citep{AveMeas}. A cosmic-ray particle passing through the instrument will traverse 16 individual tubes and generate 16 independent signals along its trajectory. The gas used in LDB2 is a 95\%:5\% Xe:CH$_{4}$ mixture at a pressure of one atmosphere.

For each cosmic-ray nucleus that traverses the TRACER instrument, the nuclear charge $Z$, and the energy $E$ (or the Lorentz-factor $\gamma=E/mc^2$), as well as the geometric trajectory through the instrument are determined. 

The charge is obtained from the signals of the \v{C}erenkov-scintillator combination,  utilizing the fact that electromagnetic interactions scale with $Z^2$. However, the energy dependence of these two signals is different: the scintillation signal decreases with $\beta^{-2}$, and the \v{C}erenkov signal rises as $(1-1/n^2\beta^2)$ above a threshold $\beta=n^{-1}$ (with refractive index n=1.49, and $\beta=v/c$).  With $\beta\sim1$, the scintillation signal becomes independent of energy  at a ``minimum ionization level.''  Thus, if the two signals are combined, the dependence on energy (or $\beta$) can be eliminated, and $Z$ can be uniquely determined within measurement fluctuations. There are two complications to these simple relations: the signal output of the scintillator does not scale exactly like $Z^2$, and contributions of signals due to $\delta$-rays lead to an additional component in the \v{C}erenkov light and generate a slight increase in the scintillation signal above the minimum ionization level. These effects have been discussed previously by~\citet{Gahbauer,AveMeas,NIM}.

\begin{figure}[tb]
 \includegraphics[width=\linewidth]{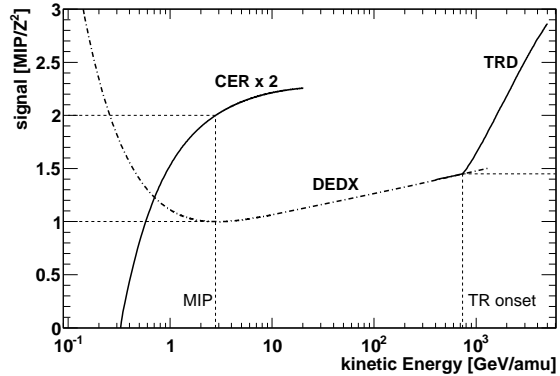}
 \caption{Response functions, in units of signal at minimum ionizing energy (MIP) normalized by $Z^2$, of sub-detectors used for energy measurements. The dashed lines indicate the signals at minimum ionizing energy and TR onset.\label{fig2}}
\end{figure}

The energy measurement of TRACER depends on the energy response of the \v{C}erenkov signal, the magnitude of the relativistic increase in specific ionization in the gas-proportional tubes, and on the energy dependence of the transition radiation (TR) signal. The characteristic response curves are shown in Figure~\ref{fig2}. The very rapid rise of the \v{C}erenkov signal permits an accurate measurement at low energies ($\sim$0.5 to 3~GeV~amu$^{-1}$). The much slower rise of the ionization signal allows for an energy determination in the range of $\sim10$ to a few hundred GeV~amu$^{-1}$, if the magnitude of this rise is larger than the level of fluctuations in the signals. Finally, the highest energies can be rather precisely determined from the rapid increase of the TR intensity. Because of the steeply falling energy spectrum of cosmic rays, the measurement at high energies is limited by counting statistics and not by saturation of the detector response.

\begin{figure}[tb]
  \includegraphics[width=\linewidth]{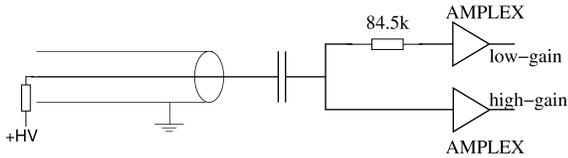}
  \caption{Schematic of the dual gain readout of the proportional tubes.\label{fig3}}
\end{figure}

For LDB2, detector upgrades were implemented in three areas to include the light elements boron ($Z=5$), carbon ($Z=6$), and nitrogen ($Z=7$) in the measurement:
\begin{enumerate}
 \item The signals from each wire of the proportional tubes were split resistively into two amplifiers (see Figure~\ref{fig3}), increasing the overall dynamic range of from $\sim10^2$ to $\sim10^4$ (i.\ e.\ charge signals from $\sim10^{-15}$ Coulomb to nearly $10^{-11}$ Coulomb). This change required a doubling of the number of amplifier channels (provided by AMPLEX VLSI-chips~\citep{amplex}), and a complete redesign of the digitization and data acquisition electronics.
 \item The xenon content in the gas mixture for the proportional tubes was increased by almost a factor of 4 as compared to LDB1. As a result, the energy resolution is considerably enhanced (see Figure~\ref{fig9} and Section~\ref{sec:Primary}). The increased xenon content also causes a slight change in the TR response. The upgrade facilitates the energy measurements for boron and carbon in the 100~GeV~amu$^{-1}$ energy range.
 \item While for LDB1 only a scintillator on top and a scintillator-\v{C}erenkov counter pair on the bottom were employed, two identical scintillator-\v{C}erenkov counter pairs were now incorporated on the top and bottom. This permits independent charge measurements on top and bottom and results in an overall improvement of the charge determination.
\end{enumerate}

\subsection{Balloon Flight} 

\begin{figure}[tb]
  \includegraphics[width=\linewidth]{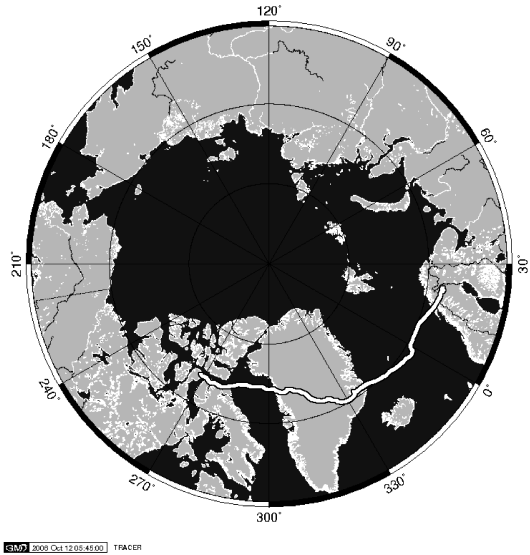}
  \caption{Flight trajectory of LDB2 from Kiruna (Sweden) to Northern Canada.\label{fig4}}
\end{figure}

\begin{figure}[tb]
 \includegraphics[width=0.75\linewidth]{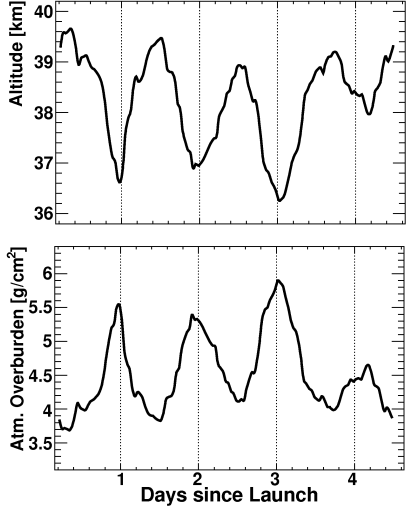}
 \caption{Profiles of altitude and vertical atmospheric overburden as a function of flight time for LDB2.\label{fig5}}
\end{figure}

The TRACER instrument was launched on a long-duration balloon flight in the Northern hemisphere from Kiruna (Sweden) on July 8, 2006. While it was intended to complete at least one circular path around the North Pole, nominally taking 14 days, the flight was terminated over Northern Canada after 5 days as permission for overflight of Russian territory could not be obtained. Figure~\ref{fig4} shows the flight trajectory. The balloon floated at an average altitude of $\sim38$ km (corresponding to 4.5~g~cm$^{-2}$ of vertical residual atmosphere), with daily variations as shown in Figure~\ref{fig5}. The instrument operated well, and $\sim3\times10^7$ cosmic-ray events were recorded. All electronics performed as expected. The proportional tube system performed without problems during the whole flight. After 1.5~days of flight, a subset of PMTs in the bottom scintillator and in the top \v{C}erenkov detector had to be deactivated due to an electric discharge. This had only a minor impact on the instrument's performance.

For each event, the signals from all PMTs are recorded. A zero-suppression system is utilized for the proportional tubes, and only signals above a threshold value are stored. All events are assembled and processed by the on-board CPU for storage on a hard disk, and are also sent to the telemetry transmitters. A total of 50~GB of raw data, at an event rate of $\sim120$~Hz, were gathered. The lifetime of the instrument was 203,200~s, at a dead time of the data acquisition of $\sim18$\%.

\section{Data Analysis}

The data analysis follows the structure of the analysis performed for the data of LDB1~\citep{AveMeas}. Events are accepted only if the following ``quality cuts'' are satisfied: (1) a minimum of three PMT's in each of the scintillation and \v{C}erenkov counters must show a non-zero signal; (2) the event  must correspond to a well reconstructed trajectory; and (3) the signal amplitudes measured in the 16 proportional tube layers must be consistent within expected levels of fluctuation. These cuts have a combined efficiency of $\sim80$\%, independent of energy (see Table~\ref{tab1}). An outline of the analysis is given in the following, with emphasis on the effect of the modifications for LDB2.

\subsection{Trajectory Reconstruction}

First, the trajectory of every cosmic-ray nucleus through the instrument is determined. A first-level reconstruction that fits straight lines to the center positions of all tubes that recorded a signal yields a lateral precision of $\sim5$~mm. A second-level fit utilizes the fact that the relative signal amplitudes are proportional to the track length in each tube, and thus are a measure for the ``impact parameters'' of the track from the tube center. The track fit then has a lateral precision of $\sim2$~mm. This corresponds to an uncertainty in total track length through the proportional tubes of $\sim3$\%.

\subsection{Charge Measurement}
\label{sec:Charge}

The charge measurement relies on the combination of the signals of the scintillator and \v{C}erenkov detectors as described previously~\citep{AveMeas,NIM}. The additional \v{C}erenkov detector used in LDB2 facilitates independent charge measurements at both the top and bottom of the instrument.

\begin{figure}[tb]
 \includegraphics[width=\linewidth]{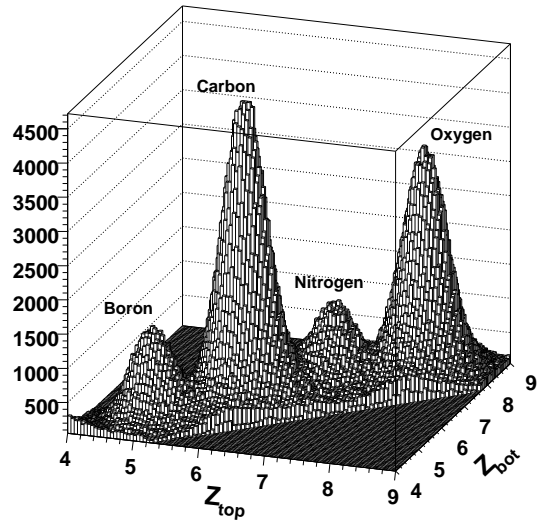}
 \caption{Two-dimensional charge distribution for light nuclei.\label{fig6}}
\end{figure}

The charges obtained for each event are shown in Figure~\ref{fig6} for the light elements boron, carbon, nitrogen, and oxygen. The elements are clearly separated with an effective charge resolution of 0.23 charge units for boron and carbon nuclei, and 0.25 charge units for nitrogen and oxygen nuclei. The resolution rises for higher charges, reaching 0.55 charge units for iron nuclei.

\begin{figure}[tb]
 \includegraphics[width=\linewidth]{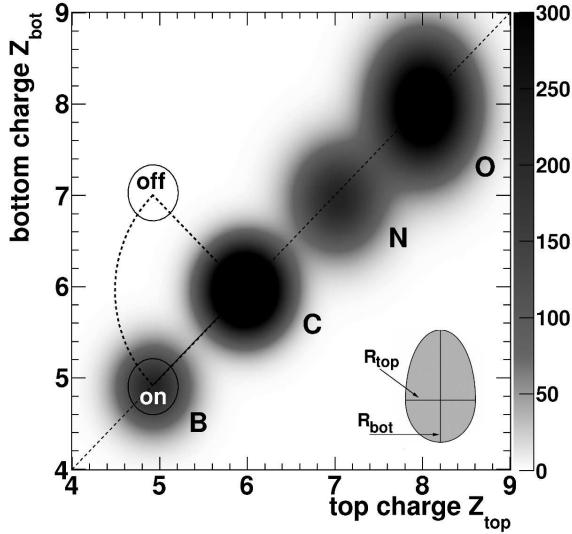}
 \caption{Parametrization of the charge distribution of top and bottom charge for light nuclei. The shape of the selection cut is indicated. The on and off region method to estimate the carbon contamination in the boron data sample is illustrated.\label{fig7}}
\end{figure}

The distributions in Figure~\ref{fig6} are parametrized with two-dimensional Gaussian functions and are shown in Figure~\ref{fig7}. Semi-elliptical cuts, commensurate with the resolution of the distributions, are applied as indicated in the figure. The cut values $R_{top}$ and $R_{bot}$ (see Figure~\ref{fig7}) are chosen to minimize the contamination by adjacent charges and are different for the individual elements. For instance, for iron nuclei $R_{top}= \pm1.0$ (in charge units) while for boron nuclei, $R_{top}= \pm0.3$. The cut $R_{bot}$ is asymmetric, with $R_{bot} = (-1.2/+2.4)$ for iron, but $R_{bot} = (-0.32/+0.64)$ for boron. The abundant elements oxygen and iron are then selected with an efficiency of 68\% (see Table~\ref{tab1}) with negligible contamination from neighboring elements. Tighter cuts applied to boron and carbon nuclei lead to efficiencies of 36\% and 44\%, respectively. Even then, the selected boron events at high energy contain a very small carbon contamination of $0.63\pm0.03$\%. This contamination level has been determined by two independent methods: (1) by moving the boron selection cut around the carbon peak to an ``off'' position (see Figure~\ref{fig7}) and counting the events within the cut; (2) by integrating the tail of the parametrized carbon distribution within the selection cut for boron. Both methods lead to the same result. There is no contamination of the boron sample by beryllium nuclei. The effect of the contamination of the boron events on the determination of the boron energy spectrum is discussed in Section~\ref{sec:Boron}.

\begin{figure}[tb]
  \includegraphics[width=\linewidth]{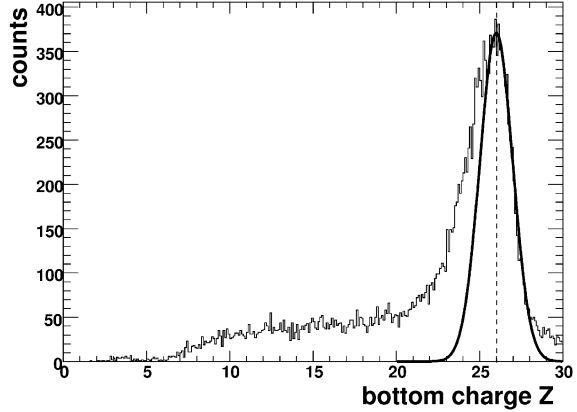}
  \caption{Bottom Charge $Z_{bot}$ distribution of events identified as iron at the top of the instrument. The solid line represents a Gaussian fit, which illustrates the distribution of non-interacting nuclei.\label{fig8}}
\end{figure}

A fraction of events undergo a charge-changing interaction within the material of the instrument (7.7~g~cm$^{-2}$). Such events exhibit an apparent lower nuclear charge at the bottom of the instrument, since the sum of the signals of the daughter nuclei is smaller than the signal of the parent nucleus. The interacting fraction can either be experimentally determined, as illustrated in Figure~\ref{fig8} for iron nuclei, or can be calculated, using the interaction cross sections in the parametrization of~\citet{BPform, Westphal}, which are assumed to be independent of energy. The results of both methods are in agreement and one obtains a survival probability for iron nuclei through the instrument of $\sim50$\%. Similarly, the Bradt and Peters formula is used to determine interaction losses in the residual atmosphere above the balloon. The survival probabilities for all elements considered in the present work are given in Table~\ref{tab1}. They are evaluated at an average angle of incidence of $\theta=30^\circ$ (see also~\citet{NIM}). The interaction losses are taken into account in the calculation of the effective aperture of TRACER for each element (see Table~\ref{tab1}).

\subsection{Energy Assignment}
\label{sec:Primary}

The energies, or the Lorentz-factors $\gamma=E/mc^2$, of the individual primary cosmic-ray nuclei (C, O, Fe) are determined by the same procedures that were employed for LDB1 and were described previously~\citep{AveMeas}.

\textit{At low energies ($\sim$0.5 to 3.0 GeV~amu$^{-1}$),} the energy measurement is performed from the signals of the bottom \v{C}erenkov counter.  The counter has a steep energy response in this region (see Figure~\ref{fig2}), hence, it exhibits excellent energy resolution.

\textit{At high energies,} the dE/DX signals from the entire proportional tube array provide the energy measurement over the range of $\sim10$ to $400$~GeV~amu$^{-1}$. The correlation between the dE/dx-array and the TRD identifies particles above $\sim1000$~GeV~amu$^{-1}$. Because of the sudden change in the response function around the onset of TR, the energy resolution is not well defined in the energy range of about 400 to 1500~GeV~amu$^{-1}$. Therefore, this energy region is excluded from analysis.

\begin{figure}[tb]
 \includegraphics[width=\linewidth]{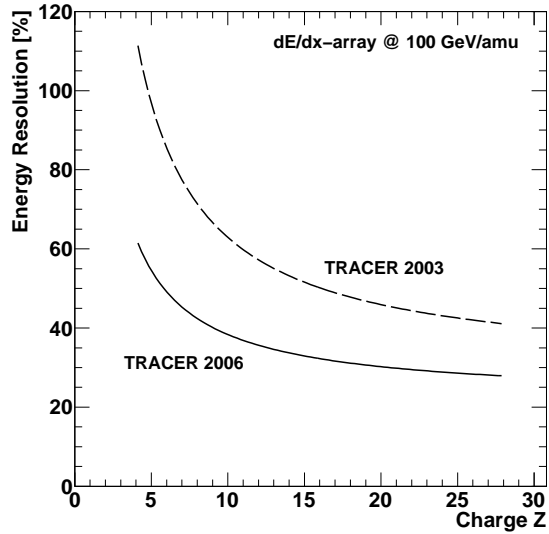}
 \caption{Measured energy resolutions (1$\sigma$) for LDB1 and LDB2 for the dE/dx-array.\label{fig9}}
\end{figure}

The increase in the xenon content of the proportional tubes modifies the detector response in two respects: 
(1) the relative magnitude of the relativistic increase in dE/dx is slightly reduced to 33\% instead of 40\% over the $\gamma$-range from 10 to 440. However, the absolute signal is increased by nearly a factor of 4. Consequently, the relative signal fluctuations are reduced by about a factor of two, and the relative energy resolution between 10 and several 100 GeV~amu$^{-1}$ is improved by the same factor. This important fact is illustrated in Figure~\ref{fig9}.
(2) for the TRD, both the signal contribution from ionization, and the conversion probability of x-rays are increased as compared to LDB1 because of the larger xenon content of the tubes. In balance, the appearance threshold for x-rays moves to a higher $\gamma$-value of $785\pm140$ and the relative magnitude of the TR signal decreases by $19\pm3$\%. More detail on these calibrations is given by~\citet{NIM} and references therein.
 
\begin{figure}[tb]
 \includegraphics[width=\linewidth]{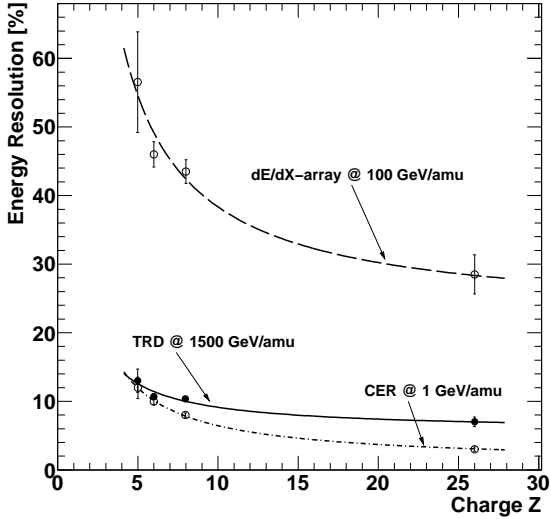}
 \caption{Measured energy resolution (1$\sigma$) of individual detector subsystems vs.\ charge $Z$ for typical energies for LDB2.\label{fig10}}
\end{figure}

The rather steep energy dependence of the TR response curve (Figure~\ref{fig2}) insures good energy resolution for the highest energy events. The measured energy resolution for all three sub-detectors is shown in Figure~\ref{fig10} as a function of charge Z.

The accepted high-energy events are characterized by two quantities, the average ionization loss $\langle dE/dx\rangle$, and the average TRD signal $\langle dE/dx +TR \rangle$, which are defined as the sums of the signals in the dE/dx-array and TRD, respectively, divided by the sum of the track lengths. It is vital for the energy assignment in the high energy region that particles at energies below the minimum ionization level are excluded as the response function (Figure~\ref{fig2}) is double-valued and rises towards lower energies. This is accomplished with high efficiency by a cut on the bottom \v{C}erenkov signal at the level corresponding to minimum ionization ($\gamma=3.97$).

\begin{figure}[tb]
 \includegraphics[width=\linewidth]{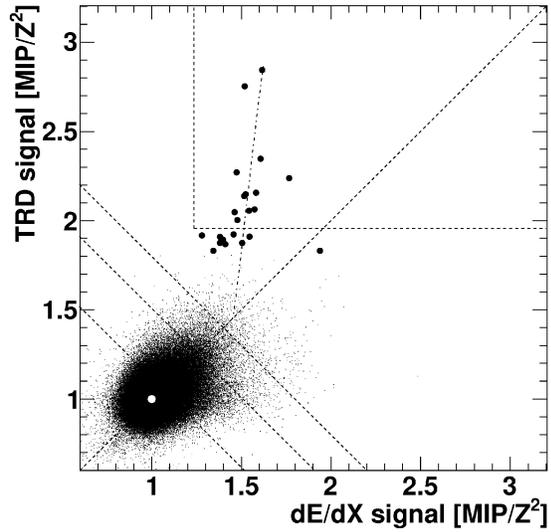}
 \caption{dE/dx-array signal vs.\ TRD signal for high-energy oxygen data. The diagonal and energy bin edges (dashed) are indicated, as well as the TR rsponse (dash-dotted). Minimum ionizing energy is indicated as a white marker. Possible TR events are shown in bold symbols.\label{fig11}}
\end{figure}

Figure~\ref{fig11} shows a cross correlation of $\langle dE/dx\rangle$ and $\langle dE/dx+TR\rangle$ for oxygen nuclei (Z=8) above 3 GeV~amu$^{-1}$. At energies below TR threshold ($\sim1.4$~MIP, where MIP is the signal of a minimum ionizing particle, at an energy of $\sim3$~GeV~amu$^{-1}$) both signals are the same within fluctuations. The distribution peaks around 1~MIP and rises along the diagonal as the energy increases. After the onset of TR, the events rise above the diagonal with increasing energy. Unambiguous TR-events, well outside of possible dE/dx fluctuations, are highlighted in the scatter plot. These events have energies well above 1,000~GeV~amu$^{-1}$. Note that their number represents less than $10^{-4}$ of the total number of events in the plot. Nevertheless, the plot is remarkably free of background and demonstrates the power of identifying particles by independent measurements of ionization loss and TRD signal. The dashed lines in the plot also indicate the edges of the energy bins into which the data are sorted. The same procedures have been followed for the primary elements iron (Z=26) and carbon (Z=6).

To generate the energy spectra of the individual elements, the events are sorted into energy bins as indicated (with a gap between about 400 to 1500~GeV~amu$^{-1}$ for reasons of energy resolution), and ``overlap corrections'' are applied to account for possible misplacements into neighboring bins. The corresponding correction factors are given in Table~\ref{tab1}. They are close to unity as the energy bins are chosen to be quite wide with respect to the energy resolution. The exception are the low energy dE/dx bins for boron and carbon, where signal fluctuations are large.

\begin{deluxetable}{ccccccc}
\tablecaption{Summary of efficiencies and exposure factors.\label{tab1}}
\tabletypesize{\small}
\tablehead{\colhead{Item} & \colhead{Spectrum Range} & \colhead{All Elements} & \colhead{Boron} & \colhead{Carbon} & \colhead{Oxygen} & \colhead{Iron}}
 
\startdata

survival probability   & all & --- & 0.73 & 0.72 & 0.69 & 0.48 \\
instrument             &     & & & & & \\ \hline

survival probability   & all & --- & 0.80 & 0.79 & 0.77 & 0.65 \\
atmosphere             &     & & & & & \\ \hline

exposure factor & all & --- & 5.22 & 5.09 & 4.72 & 2.81 \\ 
(m$^{2}$sr days)  & & & & & & \\ \hline

           & CER &0.043 & --- & --- & --- & --- \\
zenith cut & dE/dx & --- & --- & --- & --- & --- \\
           & TRD & --- & --- & --- & --- & --- \\ \hline

         & CER & --- & 0.78 & 0.79 & 0.79 & 0.80 \\
quality cuts  & dE/dx & --- & 0.79 & 0.80 & 0.81 & 0.79 \\
         & TRD & --- & 0.79 & 0.80 & 0.81 & 0.79 \\ \hline

         & CER & --- & 0.49 & 0.56 & 0.81 & 0.78 \\
charge cut & dE/dx & --- & 0.36 & 0.44 & 0.68 & 0.67 \\
         & TRD & --- & 0.36 & 0.44 & 0.68 & 0.67 \\ \hline

                  & CER & --- & 0.9  & 0.9  & 0.9  & 1.0 \\
correlation norm. & dE/dx\tablenotemark{a} & --- & 0.2/0.7    & 0.6/0.9      & 0.7/0.9      & 0.9 \\
                  & TRD & --- & 1.0          & 1.0          & 1.0          & 1.0 \\ \hline

\enddata
\tablenotetext{a}{Two individual values are given for the two dE/dx energy bins since they differ significantly.}
\end{deluxetable}

\section{Energy Spectra}
\subsection{The Primary Nuclei C, O, and Fe}
\label{sec:PrimarySpectra}

The differential intensity for each element in an energy bin $j$ is derived from the number of events $\Delta N_{j}$ in the bin of width $\Delta E_{j}$ as:
\begin{equation}
 \left( \frac{dN}{dE}\right)_{j}=\frac{1}{\Omega}\;\frac{\Delta N_{j}}{\Delta E_{j}}\;\frac{\omega_{j}}{\epsilon_{j}},
\label{eq:Flux}
\end{equation}
with the total exposure $\Omega$, the total efficiency $\epsilon_{j}$, and the overlap correction $\omega_{j}$. For a power-law spectrum with spectral index $\alpha$, the intensity is plotted at an energy $\hat E_j$, defined as \citep{lafferty}:
\begin{equation}
 \hat E_j = \left[\frac{1}{\Delta E_j}\;\frac{1}{1-\alpha}\cdot\left( E_{j+1}^{1-\alpha}-E_{j}^{1-\alpha} \right)  \right]^{-1/\alpha},
\label{eq:Eplot}
\end{equation}
with $\Delta E_j=E_{j+1}-E_{j}$. Spectral indices $\alpha$ commensurate with previous measurements are used (2.65 at high energies for primary elements). However, the value of $\hat E_j$ is not very sensitive to the choice of $\alpha$. The highest-energy data point is an integral value above $E_j$, i.\ e.\ $E_{j+1}\rightarrow\infty$. The corresponding differential intensity is then
\begin{equation}
\left(\frac{dN}{dE} \right)_{j} =\Delta N_{j}\cdot\frac{1}{\Omega}\;\frac{\omega_{j}}{\epsilon_{j}}\;\frac{(\alpha-1)\hat E_{med}^{-\alpha}}{E_{j}^{1-\alpha}},\\
\label{eq:IntInt}
\end{equation}
and is plotted at the median energy $\hat E_{med}$:
\begin{equation}
\hat E_{med}=\left[\frac{E_{j}^{1-\alpha}}{2} \right]^{1/(1-\alpha)}.
\label{eq:IntE}
\end{equation}

\begin{figure}[tb]
 \includegraphics[width=\linewidth]{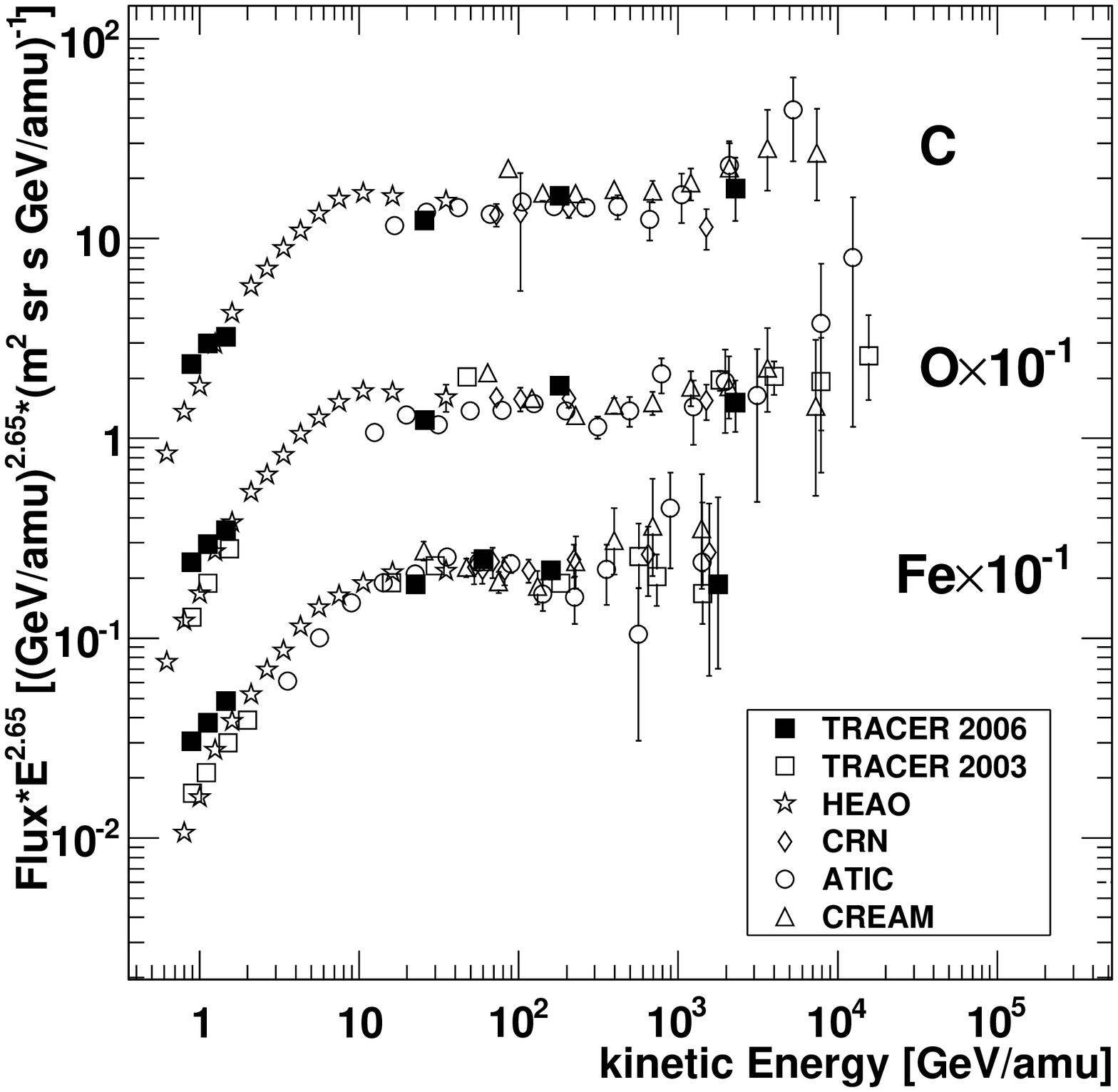}
 \caption{Differential energy spectra of carbon, oxygen, and iron (top to bottom). The spectra are multiplied by $E^{2.65}$, as well as divided by 10 for oxygen and iron. For comparison, results from HEAO~\citep{HEAObc}, CRN~\citep{crn}, ATIC~\citep{aticspectra}, and CREAM~\citep{CREAMspectra} are shown. The error bars are statistical only. \label{fig12}}
\end{figure}

The differential energy spectra for iron, oxygen, and carbon nuclei are given in terms of absolute intensities on top of the atmosphere in Figure~\ref{fig12}, and are summarized in Table~\ref{tab2}. Uncertainties are stated separately for statistical and systematic contributions. The figures include previously published data for comparison, and the energy spectra are shown multiplied by $E^{2.65}$ for clarity.

The statistical uncertainties are based on Poisson counting statistics. For ten or fewer events, errors become asymmetric and take values according to~\citet{gehrels} for a confidence level of 84\%\footnote{More recent work by~\citet{Feldman} leads to very similar values.}. The systematic errors represent uncertainties for energy bin overlaps, selection cut efficiencies, and imperfect knowledge of the response functions.

\begin{deluxetable}{cccr@{}l@{$\times$}l}

\tabletypesize{\small}
\tablecaption{Measured intensities.\label{tab2}}
\tablecolumns{6}
\tablehead{
\colhead{Energy Range} &
\colhead{Energy $\hat E$} &
\colhead{Number of Events} &
\multicolumn{3}{c}{Intensity $\pm\sigma_{stat}\pm\sigma_{sys}$} \\
\colhead{GeV~amu$^{-1}$} &
\colhead{GeV~amu$^{-1}$} &
\colhead{} &
\multicolumn{3}{c}{(m$^2$ sr s GeV~amu$^{-1}$)$^{-1}$}
}
\startdata
\sidehead{Iron (Z=26)}
0.8 -- 1.0 & 0.9 & 593 & (4.1 $\;$&$\pm$ 0.2 $\pm$ 0.1) &$\;10^{-1}$   \\
1.0 -- 1.3 & 1.1 & 487 & (2.78 $\;$ &$\pm$ 0.1 $\pm$ 0.09) &$\;10^{-1}$   \\
1.3 -- 1.7 & 1.5 & 386 & (1.79 $\;$ &$\pm$ 0.07 $\pm$ 0.2) &$\;10^{-1}$   \\
14 -- 38 & 23 & 1180 & (4.8 $\;$ &$\pm$ 0.1 $\pm$ 1.3) &$\;10^{-4}$   \\
38 -- 101 & 60 & 314 & (4.7 $\;$ &$\pm$ 0.2 $\pm$ 1.7) &$\;10^{-5}$   \\
101 -- 264 & 159 & 57 & (3.2 $\;$ &$\pm$ 0.3 $\pm$ 1.4) &$\;10^{-6}$   \\
1200 --  & 1800 & 1 & (5 $\;$ &$\pm^{10}_{4}\,\pm^{2}_{3}$) &$\;10^{-9}$   \\
  
\sidehead{Oxygen (Z=8)}
0.8 -- 1.0 & 0.9 & 7900 & (3.26 $\;$&$\pm$ 0.04 $\pm$ 0.09) &$\;10^0$   \\
1.0 -- 1.3 & 1.1 & 6440 & (2.17 $\;$ &$\pm$ 0.03 $\pm$ 0.05) &$\;10^0$   \\
1.3 -- 1.7 & 1.5 & 6210 & (1.28 $\;$ &$\pm$ 0.02 $\pm$ 0.05) &$\;10^{0}$   \\
9.5 -- 87 & 26 & 43700 & (2.25 $\;$ &$\pm$ 0.01 $\pm$ 0.06) &$\;10^{-3}$   \\
87 -- 432 & 181 & 2140 & (1.91 $\;$ &$\pm$ 0.04 $\pm$ $^{0.6}_{0.8}$) &$\;10^{-5}$   \\
1500 -- & 2300 & 12 & (1.9 $\;$ &$\pm$ 0.6 $\pm^{0.8}_{1.1}$) &$\;10^{-8}$   \\
 
\sidehead{Carbon (Z=6)}
0.8 -- 1.0 & 0.9 & 6510 & (3.19 $\;$ &$\pm$ 0.04 $\pm$ 0.1) &$\;10^0$   \\
1.0 -- 1.3 & 1.1 & 5460 & (2.19 $\;$ &$\pm$ 0.03 $\pm$ 0.1) &$\;10^0$   \\
1.3 -- 1.7 & 1.5 & 4760 & (1.19 $\;$ &$\pm$ 0.02 $\pm$ 0.1) &$\;10^{0}$   \\
9.5 -- 87 & 26 & 28300 & (2.17 $\;$ &$\pm$ 0.01 $\pm$ 0.06) &$\;10^{-3}$   \\
87 -- 432 & 181 & 1430 & (1.70 $\;$ &$\pm$ 0.04 $\pm$ $^{0.5}_{0.7}$) &$\;10^{-5}$   \\
1500 -- & 2300 & 10 & (2.2 $\;$ &$\pm^{0.9}_{0.7}\,\pm^{0.9}_{1.3}$)&$\;10^{-8}$   \\ 

\sidehead{Boron\tablenotemark{a} (Z=5)}
0.8 -- 1.0 & 0.9 & 2170 & (1.13 $\;$&$\pm$ 0.03 $\pm$ 0.1)&$\;10^0$   \\
1.0 -- 1.3 & 1.1 & 1600 & (7.3 $\;$&$\pm$ 0.2 $\pm$ 0.1) &$\;10^{-1}$   \\ 
1.3 -- 1.7 & 1.5 & 1280 & (3.8 $\;$&$\pm$ 0.1 $\pm$ 0.1) &$\;10^{-1}$   \\
9.5 -- 86.5 & 23.8 & 7200 (27) & (4.9 $\;$&$\pm$ 0.1 $\pm$ 1.0) &$\;10^{-4}$   \\
86.5 -- 432 & 173 & 413 (48) & (1.6 $\;$&$\pm$ 0.1 $\pm$ 0.4) &$\;10^{-6}$   \\
1500 -- & 2070 & 5 (4) & (4 $\;$&$\pm^{8}_{3}\,\pm$ 2) &$\;10^{-9}$   \\ 

\enddata
\tablenotetext{a}{in brackets: expected number of carbon background}
\end{deluxetable}

\subsection{Energy Spectrum of Boron}
\label{sec:Boron}

Boron is a secondary element produced by spallation mainly of carbon and oxygen. These parent nuclei are much more abundant than boron nuclei. Therefore, two additional considerations must be taken into account when deriving the energy spectrum of boron nuclei: (1) contamination of the boron sample with a small admixture of carbon nuclei (see Section~\ref{sec:Charge}), and (2) the contribution of boron nuclei generated in the residual atmosphere above the balloon.

\begin{figure}[tb]
 \includegraphics[width=\linewidth]{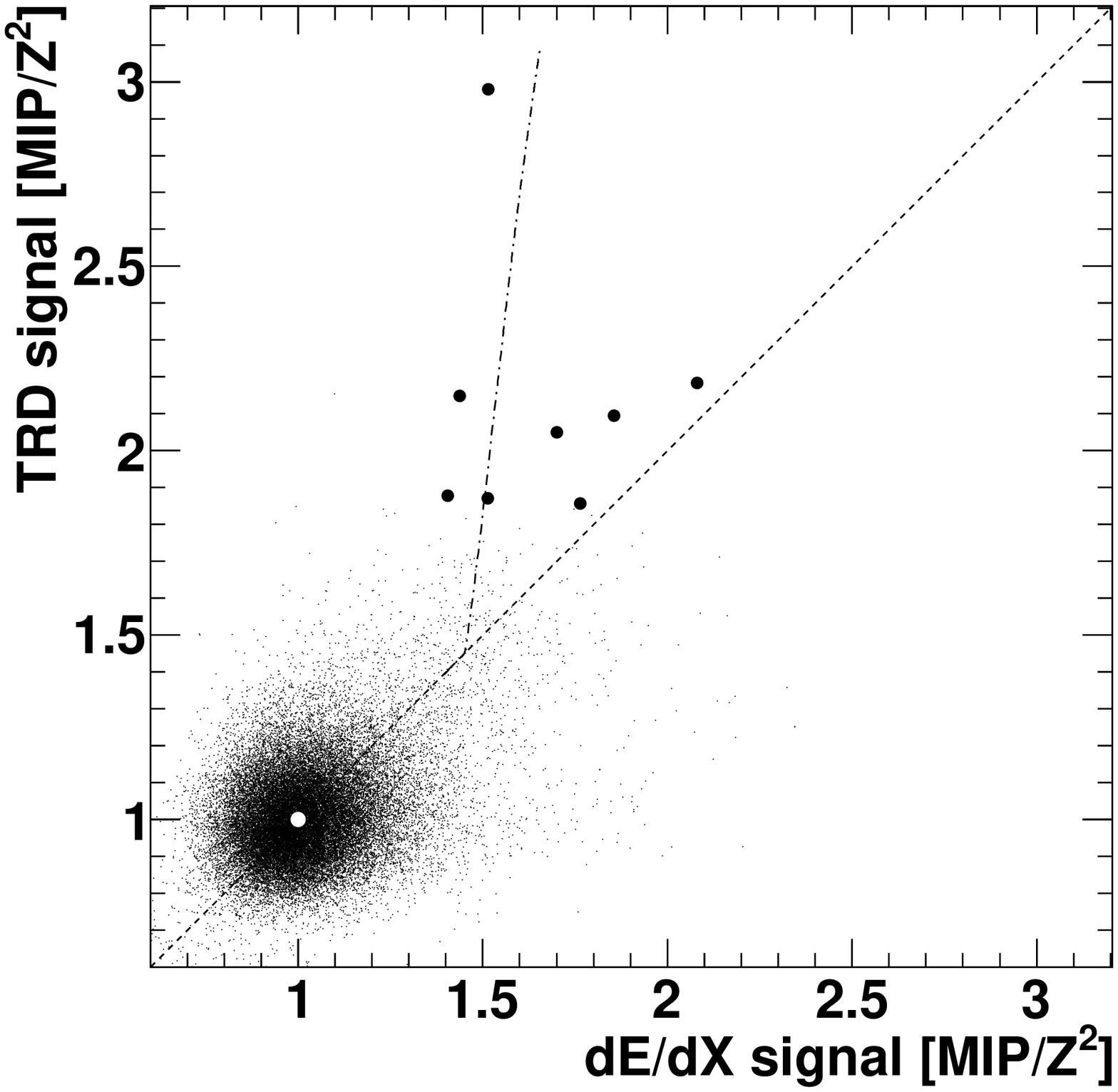}
 \caption{dE/dx-array signal vs.\ TRD signal for high energy boron data. Small and thick black markers represent events at low and high energy, respectively. Minimum ionizing energy is indicated as a white marker. The diagonal is shown as a dashed line, and the dash-dotted line represents the TR response.\label{fig13}}
\end{figure}

\begin{figure}[tb]
 \includegraphics[width=\linewidth]{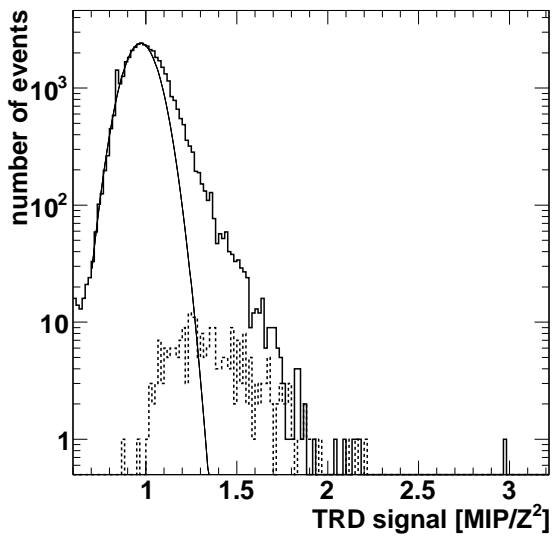}
 \caption{Histogram of TRD signals. The solid histogram represents boron data. The thin line indicates the distribution of minimum ionizing nuclei. The dashed histogram shows a distribution of contaminating carbon events as expected from simulation. Note that the vertical axis is logarithmic.\label{fig14}}
\end{figure}

To investigate the carbon contamination, we consider the scatter plot $\langle dE/dx+TR\rangle$ vs. $\langle dE/dx\rangle$ shown in Figure~\ref{fig13}. This plot is derived from about 40,000 boron nuclei. It shows qualitatively the same behavior as Figure~\ref{fig11} for oxygen, although the relative statistical fluctuations are larger due to the lower Z of boron. However, an analysis of the charge selection for boron (see Section~\ref{sec:Charge}) based on Figures~\ref{fig6} and \ref{fig7}, predicts that Figure~\ref{fig13} must include a contamination of $250\pm12$ carbon nuclei ($0.63\pm0.03$\%), clustered around the MIP signal of carbon events. The effect of this background is illustrated in the histogram of $\langle dE/dx +TR\rangle$ for boron in Figure~\ref{fig14}, which represents the projection of the scatter plot of Figure~\ref{fig13} upon the vertical axis. As expected, the histogram is asymmetric towards higher signals due to the relativistic rise and due to the appearance of TR photons. Also included in Figure~\ref{fig14} is a simulated distribution of the 250 carbon background events. This  background resides mostly in the region of tail of the boron distribution, peaking near the onset of TR. Obviously, the carbon contamination does not affect the boron measurement at low energy, below $\sim100$~GeV~amu$^{-1}$. However, the contamination makes the identification of boron unreliable in the region $\sim450-1500$~GeV~amu$^{-1}$, which is already excluded from analysis due to the ill-defined energy resolution around the onset of TR. There still remains at least one event at very high energy which very unlikely (at a probability of $10^{-5}$) is a carbon nucleus. This event also clearly stands out in the scatter plot of Figure~\ref{fig13} and appears to be a genuine boron nucleus of $\sim6,000$~GeV~amu$^{-1}$.

The second effect to be accounted for is the generation of secondary boron nuclei by spallation of heavier nuclei in the residual atmosphere above the balloon. Each galactic component is attenuated in the atmosphere, at an average column density\footnote{Averaged over the duration of the flight and over the zenith-angle distribution of accepted particles.} $X=5.2$~g~cm$^{-2}$, by a factor $\exp(-X/\Lambda)$. Here, $\Lambda=m/\sigma$ is the attenuation path length, with $m$ the average mass of the target atoms in air, and $\sigma$ the spallation cross section. This attenuation (see also Section~\ref{sec:Charge}) has been taken into account in deriving the energy spectra of the primary elements (Section~\ref{sec:PrimarySpectra}).

The contribution of spallation products to the observed particle flux must be included for rare elements such as boron: the measured intensity $N_B(X)$ is then
\begin{equation}
N_B(X) =  N_B(0)\cdot\exp(-X/\Lambda_B) + \sum N_P(0) X/\Lambda_{P-B}
\label{eq:Bprod}
\end{equation}
Here, $N_P$ refers to parent nuclei heavier than boron, and $\Lambda_{P-B}$ is the differential spallation path length in air for an interaction of the parent leading to boron nuclei. To obtain $N_B(0)$, the boron intensity at the top of the atmosphere, this expression has been evaluated using differential cross sections from~\citet{webber}. Contributions to the sum in Eq.~(\ref{eq:Bprod}) come essentially only from carbon and oxygen, and, based on the results described in Section~\ref{sec:Ratios}, the carbon-to-oxygen abundance ratio is taken to be independent of energy.

An independent technique to determine the atmospheric boron contribution utilizes the daily variations of atmospheric depth during the flight (see Figure~\ref{fig5}). The corresponding variations in the counting rates for the individual elements are a direct measure of the atmospheric effects. The analysis leads to results which are consistent with those obtained with published cross sections.  More detail on this procedure will be given in a separate publication~\citep{GCurves}.

\begin{figure}[tb]
 \includegraphics[width=\linewidth]{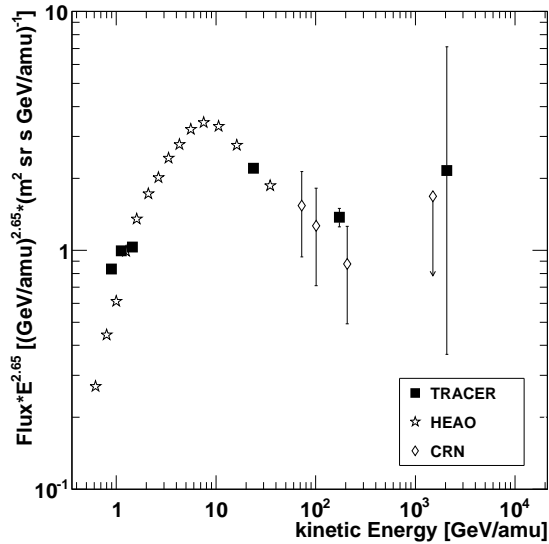}
 \caption{Differential energy spectrum of boron. The spectrum is multiplied by $E^{2.65}$. For comparison, results from HEAO~\citep{HEAObc} and CRN~\citep{CRNbc} are shown. The error bars are statistical only. \label{fig15}}
\end{figure}

The contribution of atmospheric secondary boron nuclei to the measured energy spectrum is small at low energy ($\sim6$\% at 1~GeV~amu$^{-1}$), but reaches $\sim20$\% above 1~TeV~amu$^{-1}$. Thus, the single boron event defining the highest-energy data point has a 20\% chance of being of atmospheric origin. The energy spectrum of boron after corrections for atmospheric contributions and interaction losses in both, atmosphere and instrument, is shown in Figure~\ref{fig15} and Table~\ref{tab2}.

\subsection{Relative Intensities of Primary and Secondary Cosmic Rays}
\label{sec:Ratios}

\begin{figure}[tb]
 \includegraphics[width=\linewidth]{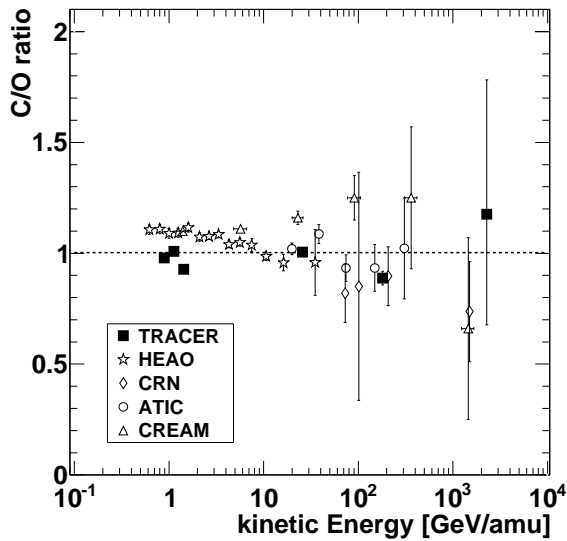}
 \caption{The carbon-to-oxygen abundance ratio as a function of kinetic energy per nucleon. Error bars are statistical only. Previous results from HEAO~\citep{HEAObc}, CRN~\citep{CRNbc}, ATIC~\citep{ATICbc}, and CREAM~\citep{CREAMbc} are shown for comparison. \label{fig16}}
\end{figure}

It was observed in LDB1 that the primary spectra of the primary cosmic-ray nuclei from oxygen to iron above 20~GeV~amu$^{-1}$ can be described by a power law in energy with a common spectral index of $2.67\pm0.05$. Consequently, the relative abundances of the different elemental species do not change with energy. The same behavior characterizes the results of LDB2, including the primary element carbon. This is illustrated in Figure~\ref{fig16} which shows the abundance ratio carbon to oxygen as a function of energy. This figure also includes previous results from other experiments.

\begin{figure}[tb]
 \includegraphics[width=\linewidth]{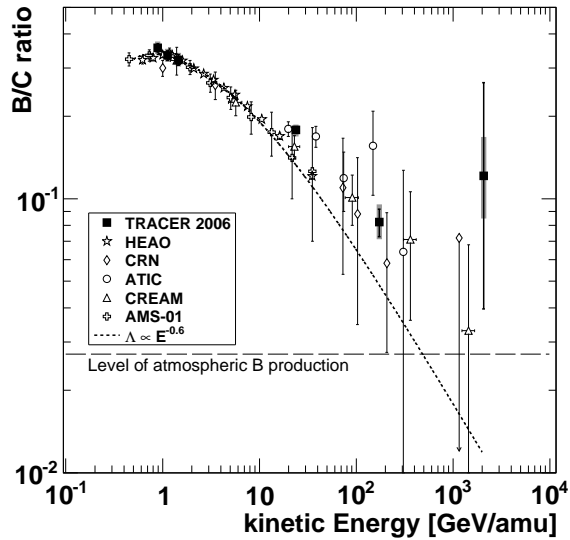}
 \caption{The boron-to-carbon abundance ratio as a function of kinetic energy per nucleon. Error bars are statistical (thin) and systematic (thick). A model corresponding to an escape path length $\propto E^{-0.6}$ (dotted) is shown. The level of atmospheric production of boron (dashed) is indicated. Previous measurements are shown from HEAO~\citep{HEAObc}, CRN~\citep{CRNbc}, ATIC~\citep{ATICbc}, CREAM~\citep{CREAMbc} and AMS-01~\citep{ams}. \label{fig17}}
\end{figure}

The relative abundance of boron, which is of prime importance for the understanding of Galactic propagation of cosmic rays, is expressed by the boron-to-carbon $(B/C)$ abundance ratio. This ratio is calculated from the differential intensities in Table~\ref{tab2}, and presented in Table~\ref{tab3} and Figure~\ref{fig17} as a function of kinetic energy per nucleon. While the same energy intervals have been used to derive the energy spectra, the mean energy in each bin differs slightly because of the different spectral slopes (Equations~(\ref{eq:Eplot}) and (\ref{eq:IntE})). For calculating the ratio, the carbon flux has been scaled to the same energies as are used for the boron spectrum. The table states both statistical and systematic uncertainties. In the figure, statistical and systematical uncertainties are shown as thin and thick error bars, respectively.

The statistical uncertainty of the boron-to-carbon ratio is derived with a procedure based on Bayes' Theorem~\citep{paterno}, which extends Poisson statistics to small numbers and small ratios. The dashed line in Figure~\ref{fig17} indicates the expected level of the boron-to-carbon ratio due to atmospheric production of boron nuclei, which has been subtracted.

Many systematic uncertainties, i.e.\ for aperture, efficiencies, and detector response functions, are correlated for the measurement boron and carbon nuclei, and thus cancel for the ratio. The main source of systematic uncertainty remaining comes from the calculation of the amount of boron nuclei produced in the atmosphere. Also, the uncertainty of the energy measurement does not entirely cancel due to the different spectral indices of the boron and carbon energy spectra.

\begin{deluxetable}{ccl} 

\tablecaption{The boron-to-carbon ratio.\label{tab3}}
\tabletypesize{\small}
\tablehead{
\colhead{Energy Range} &
\colhead{Energy $\hat E$} &
\colhead{B/C ratio $\pm\sigma_{stat}\pm\sigma_{sys}$} \\
\colhead{GeV~amu$^{-1}$} &
\colhead{GeV~amu$^{-1}$} &
\colhead{}
}

\startdata

0.8 -- 1.0 & 0.9 & ( 3.6 $\pm$ 0.1 $\pm^{0.2}_{0.2}$ )$\,\times\,10^{-1}$ \\
1.0 -- 1.3 & 1.1 & ( 3.3 $\pm$ 0.1 $\pm^{0.2}_{0.2}$ )$\,\times\,10^{-1}$ \\ 
1.3 -- 1.7 & 1.5 & ( 3.2 $\pm$ 0.1 $\pm^{0.2}_{0.2}$ )$\,\times\,10^{-1}$ \\
9.5 -- 86.5 & 23.8 & ( 1.80 $\pm$ 0.04 $\pm^{0.08}_{0.09}$ )$\,\times\,10^{-1}$ \\
86.5 -- 432 & 173 & ( 8.2 $\pm$ 0.9 $\pm^{1.3}_{1.1}$ )$\,\times\,10^{-2}$ \\
1500 -- & 2070 & ( 1.2 $\pm^{1.4}_{0.8}\pm^{0.5}_{0.4}$ )$\,\times\,10^{-1}$ \\

\enddata
\end{deluxetable}

\section{Discussion and Conclusions}

We have described a new measurement of the energy spectra of highly relativistic cosmic-ray nuclei. The TRACER instrument had previously (LDB1) provided measurements of the heavier nuclei from oxygen to iron ($Z=8$ to $26$). The most recent flight (LDB2 in 2006) includes the lighter nuclei and covers the region from boron to iron ($Z=5$ to $26$). LDB1 had a longer duration, and therefore better statistics than LDB2. Nevertheless, the two measurements show absolute intensities that agree with each other over the wide range of energy covered. This provides confidence that the new results for the elements carbon and boron (Figures~\ref{fig14} and \ref{fig15}) are consistent with the previous work.

\begin{figure}[tb]
 \includegraphics[width=\linewidth]{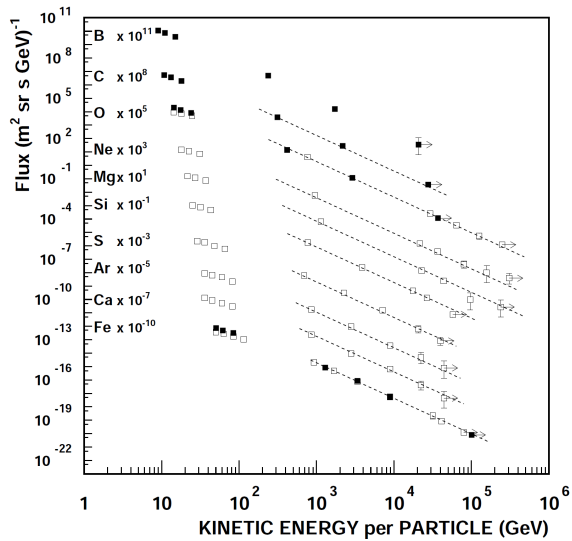}
 \caption{Compilation of the differential energy spectra measured by TRACER in LDB1 and LDB2 (see \citet{AveMeas} and this work). The dashed lines represent a simple power-law fit above 20~GeV~amu$^{-1}$.\label{fig18}}
\end{figure}

It should be noted, however, that the differential intensities from the present work at low energy (at about 1~GeV~amu$^{-1}$) are consistently higher compared to previous measurements. This can be attributed to the different solar modulation of low-energy cosmic rays~\citep{goldstein,velinov} at the time of the flight. Where overlap exists, the TRACER results are also in good agreement with measurements from HEAO~\citep{HEAObc}, CRN~\citep{crn}, ATIC~\citep{aticspectra}, and CREAM~\citep{CREAMspectra}, but have better statistics than some of these measurements. Currently, the analysis of the TRACER LDB2 data is still ongoing, concentrating on the spectra of the primary elements neon ($Z=10$), magnesium ($Z=12$), and silicon ($Z=14$). Results on these elements will be published in due course. A compilation of all present results from the two LDB flights of TRACER is given in Figure~\ref{fig18}. 

For boron, there are only two measurements besides TRACER that report absolute intensities at relativistic energies: HEAO~\citep{HEAObc} and CRN~\citep{CRNbc}, but none of these extends into the TeV~amu$^{-1}$ region. These data are included in Figure~\ref{fig15}. Good agreement with the present work can be noted within the quoted error limits, and in the energy region where overlap exists.

Most important is the energy dependence of the relative boron intensity, i.\ e.\ of the $B/C$ ratio. In Figure~\ref{fig17}, the TRACER results are compared to previous work, including the results from ATIC~\citep{ATICbc}, CREAM~\citep{CREAMbc}, and AMS-01~\citep{ams}, for which the absolute energy spectrum of boron has not been reported. Figure~\ref{fig17} also includes an extrapolation (dotted line) for an $E^{-0.6}$ decrease of this ratio with energy on the basis of measurements at lower energy~\citep{yanasak,HEAObc,CRNbc}. However, there is no model of propagation in the ISM that would predict such a behavior a priori. Furthermore, this energy dependence would imply unreasonably small values for the propagation path length of cosmic rays at the highest energies (see, e.g.~\citet{AveInt}). Thus, it is quite interesting that the data points of TRACER on the B/C ratio above 100~GeV~amu$^{-1}$, and also, as shown in Figure~\ref{fig17}, some of the results of other measurements, appear to lie above the $E^{-0.6}$ prediction. This feature may have profound implications upon our understanding of cosmic-ray propagation. It may suggest that the energy dependence of the propagation path length flattens at high energy, and perhaps indicate an asymptotic transition to a constant residual path length. A more detailed analysis of this situation will be presented in a separate publication~\citep{BCpaper}.

In conclusion, TRACER has proven to be an excellent detector for measuring the B/C ratio in the TeV~amu$^{-1}$ region. The instrument has outstanding discrimination power to identify boron nuclei at the highest energies, i.e.\ above $\sim1$~TeV~amu$^{-1}$, where contamination with carbon nuclei is no longer a concern. A 30-day flight could yield $\sim10$ boron events above 1~TeV~amu$^{-1}$ and thus, provide a definitive measurement of the propagation path length at those energies. 

\acknowledgments
We are grateful to the services of the technical staff at the University of Chicago, in particular the contributions of G.\ Kelderhouse, R.\ Northrop, D.\ Plitt, C.\ Smith, and P.\ Waltz. We thank Dr.\ M.\ Ichimura for help with the data analysis, and we acknowledge the dedicated work of a group of undergraduate students led by E.\ Brannon, and supported by the Illinois Space Grant. We are indebted to the staff of the Columbia Scientific Balloon Facility and the Esrange Space Center (Sweden) for the successful balloon flight operations. AO acknowledges support of FOM in the Netherlands (``Stichting voor Fundamenteel Onderzoek der Materie''). This work was supported by NASA through grants NNG 04WC08G, NNG 06WC05G, and NNX 08AC41G.

\end{document}